\documentclass[twoside]{LCWS11}
\usepackage[latin1]{inputenc}
\usepackage[dvips]{graphicx,epsfig,color}
\usepackage{wrapfig,rotating}
\usepackage{amssymb,amsmath,array}
\usepackage{cite}
\begin{document}
\title{
%%%%   Paper title goes here  %%%%%%%%%%%%%%
%\LaTeX\ Template for LCWS11 Proceedings} %% 
R\&D Status of FPCCD Vertex Detector for ILD }
%***********************************************************************
% AUTHORS INFORMATION AREA
%***********************************************************************
\author{Yasuhiro Sugimoto$^1$, Hirokazu Ikeda$^2$, 
Daisuke Kamai$^3$,\\
Eriko Kato$^3$, Akiya Miyamoto$^1$, 
Hisao Sato$^4$,
and Hitoshi Yamamoto$^3$
% Optional short acknowledgment: remove next line if non-needed
%\thanks{This is an optional funding source acknowledgment.}
% DO NOT MODIFY THE FOLLOWING '\vspace' ARGUMENT
\vspace{.3cm}\\
% Addresses and institutions (remove "1- " in case of a single institution)
1- High Energy Accelerator Research Organization (KEK) \\
Tsukuba, Ibaraki 305-0801, Japan
%% Remove the next three lines in case of a single institution
\vspace{.1cm}\\
2- JAXA, Japan Aerospace Exploration Agency \\
Sagamihara, Kanagawa 252-5210, Japan
\vspace{.1cm}\\
3- Department of Physics, Tohoku University \\
Sendai, Miyagi 980-8578, Japan 
\vspace{.1cm}\\
4- Department of Physics, Shinshu University \\
Matsumoto, Nagano 390-8621, Japan
}
%%***********************************************************************
% END OF AUTHORS INFORMATION AREA
%***********************************************************************

\maketitle

\begin{abstract}
Fine pixel CCD (FPCCD) is one of the candidate sensor
technologies for the vertex detector used for experiments
at the International Linear Collider (ILC). 
A vertex detector system for the International Large
Detector (ILD) using FPCCD sensors has been designed. 
We report on the R\&D status of FPCCD sensors, as well as
the vertex detector design for ILD.
\end{abstract}

\section{Introduction}
Performance goal of vertex detectors for ILC experiment
is not easy to accomplish. The requirement for the 
impact parameter resolution is better than 
$5 \oplus 10/(p\beta \sin{^{3/2}\theta})\ \mu$m.
The innermost layer should have tolerance against 
beam background of less than few \% pixel occupancy
and radiation immunity from $1\times 10^{12}$ electron$/$cm$^2 /$y.
Bunched beam of ILC could induce RF noise, which was 
a serious problem at SLD.

There are many sensor technology options proposed
to meet these challenges. One possible solution
for these issues is the fine pixel CCD (FPCCD) 
option~\cite{sugimoto05}.
FPCCD sensors will have the pixel size of $\sim 5\ \mu$m
and will be read out in $\sim 200$~ms between bunch trains.
Because there is no beam crossing during the readout,
FPCCD sensor option is completely free from the RF noise.
The present goals of the FPCCD sensor R\&D are
listed in Table~\ref{tab:rdgoal}. 
\begin{table}
\centerline{\begin{tabular}{|l|r|}
\hline
Pixel size  & 6 $\mu$m \\ \hline
Chip size  & 1 cm $\times$ 6.5 cm  \\ \hline
Readout speed & $> 10$ Mpix/s \\ \hline
Full well capacity &  $> 10000$ electrons \\ \hline
Power consumption & $< 10$ mW/ch \\ \hline
Radiation tolerance & $> 1\times 10^{13}$ electron/cm$^2$ \\
\hline
\end{tabular}}
\caption{FPCCD sensor R\&D goal.}
\label{tab:rdgoal}
\end{table}

In addition to the sensor R\&D, we need a realistic
design of the vertex detector which can be used for realistic
simulation study of the performance of the detector system
for the ILC experiment. In this report, we also discuss  engineering
R\&D and a design of the vertex detector for 
ILD (International Large Detector)~\cite{abe} which is one of the 
validated detector systems for ILC.

Our R\&D activity also includes R\&D of the readout ASIC
for FPCCD sensors.
It is however described elsewhere in these proceedings~\cite{kato}.

\section{FPCCD sensor R\&D}
Just after the decision of the RF technology for 
the ILC main linac by  the International 
Technology Recommendation Panel (ITRP) in 2004,
we have started R\&D on the FPCCD sensors
in cooperation with Hamamatsu Photonics.
We first developed fully depleted CCD sensors
with a standard pixel size of 24~$\mu$m.
Because signal charge in a pixel should not
spread over several pixels by diffusion 
in order to suppress the pixel occupancy, 
the fully depleted epitaxial layer is
essential for the FPCCD sensors.
After establishing the technology of fully-depleted
CCD sensors~\cite{sugimoto07}, prototype
FPCCD sensors have been manufactured 4 times. 
First 2 prototypes
have 12~$\mu$m pixels~\cite{sugimoto09} and latter 2 prototypes
have four different pixel sizes,12, 9.6, 8, and 6~$\mu$m
in one chip~\cite{sugimoto10}. The size of the image area of these
prototypes is about 6~mm square.

In the third prototype sensors, the 12, 9.6, and
8~$\mu$m pixels worked properly, but the 6~$\mu$m
pixels did not work at all.
In the latest prototype (the fourth prototype) sensors,
the potential profile has been improved based on
the device simulation. About a factor 2 improvement
of the depth of the potential well is expected.
By this improvement, some signal can be seen for
the 6~$\mu$m pixels
in the 4th prototype, but the signal from the pixels
far from the readout node is quite small, and we 
need still more improvement for the 6~$\mu$m pixels.

The next prototype sensors have been designed
and will be delivered in Japanese fiscal year of 2012.
This prototype is the first large size 
($\sim 1$~cm $\times 6$~cm) prototype.
It has 8 sections and each section has the pixel size
of 12, 8, or 6~$\mu$m. Horizontal shift registers
for 6~$\mu$m pixels have a size of $6\ \mu$m $\times
12\ \mu$m to improve the full-well capacity and
to decrease  resistance of the gate lines.
There are multiple wires for gate lines and there
are many bonding pads for these wires along the
longer side of the chip.

\section{Engineering R\&D}
\subsection{Sensor thinning}
It is crucial to minimize multiple scattering
of charged particles by the sensors and the
support structure (ladders) in order to achieve 
the excellent  impact parameter resolution mentioned in 
section 1.
Therefore
very low material budget is required for the 
ladders of the vertex detector in ILC experiment.
Our ultimate goal is the radiation length of
0.1\%$X_0$/layer.  

%\begin{wrapfigure}{r}{0.45\columnwidth}
\begin{figure}
\centerline{\includegraphics[width=0.6\columnwidth]{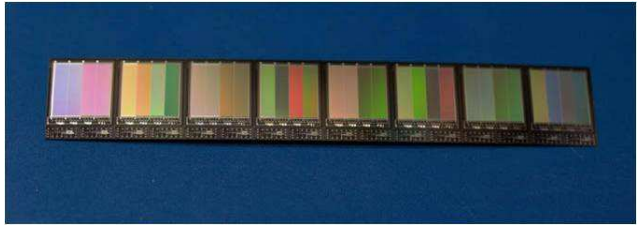}}
\caption{Prototype CCD chip thinned down to 50~$\mu$m.}
\label{fig:thin}
%\end{wrapfigure}
\end{figure}
Sensor thinning is inevitable to achieve 
the ultra light ladder. Hamamatsu Photonics 
has fabricated sample sensors thinned down to
50~$\mu$m. Figure~\ref{fig:thin} shows
the prototype CCD sensor
which has the size of 60~mm$\times$
9.7~mm$\times$50~$\mu$m. 
The wafer was thinned by mechanical grinding
after the processing
and diced into chips.
No increase of dark current was observed in these
thinned chips. Such thin sensors cannot be completely
flat. There are $\sim$0.3--0.7~mm sag over the length of 60~mm.

\subsection{Two-phase CO$_2$ cooling}
FPCCD sensors will be operated at low temperature
($\sim 40^\circ$C) inside a cryostat in order to minimize the 
effect of radiation damage, particularly minimize
the charge transfer inefficiency (CTI). 
On the other hand, the expected power consumption inside
the cryostat including sensors and readout ASICs
is close to 100~W. 
If we try to cool this vertex detector with
cold air or nitrogen gas, the flow rate will be 
quite large and could cause vibration of 
the ladders. We thus consider a cooling system
using two-phase CO$_2$.

\begin{wrapfigure}{r}{0.5\columnwidth}
%\begin{figure}
\centerline{\includegraphics[width=0.4\columnwidth]{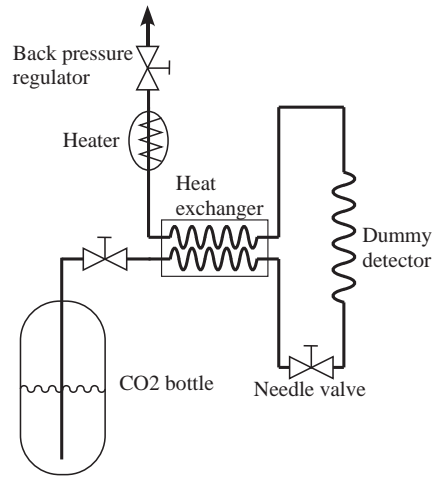}}
\caption{Schematic diagram of the CO$_2$ blow system}
\label{fig:co2}
\end{wrapfigure}
%\end{figure}
The two-phase CO$_2$ cooling system is a very 
attractive alternative because of its large 
cooling capacity of $\sim 300$~J/g.
Cooling tube will be attached to the end plates
of the vertex detector and the heat produced
by CCD output amplifiers and readout ASICs
will be removed by conduction through carbon
fiber reinforced plastic (CFRP) sheet 
put on the ladders. The return line of CO$_2$
will be used to cool the electronics (clock
drivers and data compression circuits)
placed outside the cryostat. Power consumption
of the electronics outside the cryostat is
estimated to be about 200~W/side.
The vertex detector and the inner silicon trackers
of ILD are supported by the inner support tube.
This inner support tube should be air tight
and filled with dry air in order to
prevent condensation on the cooling tube.

We have formed ``CO2 collaboration'' in Japan,
which consists of members from ILC FPCCD vertex 
detector group, ILC TPC group, Belle-II vertex 
detector group, and KEK cryogenic group.
We have constructed a blow system shown in 
Figure~\ref{fig:co2} and the
temperature of the dummy detector was 
successfully controlled between
$-40$ and $+15^\circ$C.
This system will be used for the study
of cooling of each detector.

\subsection{Common mechanical design for ILD}
In order to make a reliable detector simulation
for evaluation of the performance of the ILD,
we have to make a realistic design of the
sub-detectors. Concerning the vertex detector
in ILD, there are several options for the sensors~\cite{abe}, 
such as CMOS sensors,
DEPFET sensors, as well as FPCCD sensors.
All sensor options, however, have the same sensor
thickness of 50~$\mu$m and the same configuration
can be used for the mass production of the
simulation data. Detailed difference between
sensor options such as pixel size or point resolution
can be taken into account at the digitization and
reconstruction phase. Based on this scheme, we
decided to make a mechanical design of the 
ILD vertex detector to be used in ILD detector
simulation model (MOKKA) common to all sensor options.

\begin{figure}
\centerline{\includegraphics[width=0.95\columnwidth]{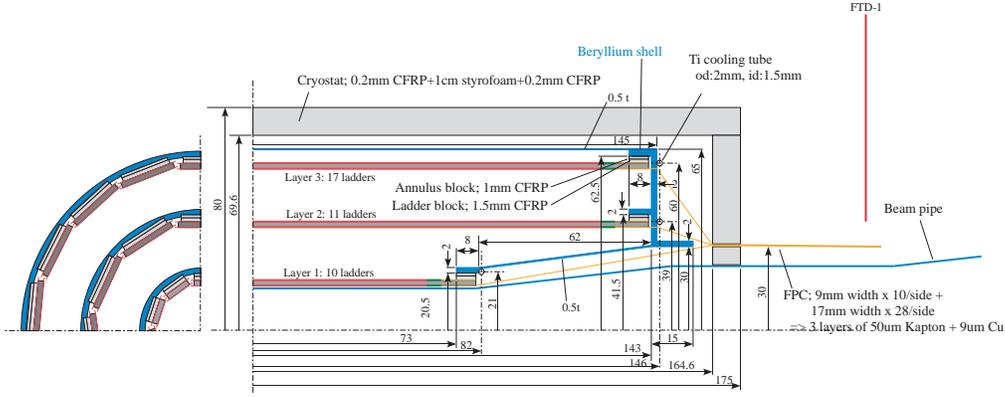}}
\caption{Common mechanical design of ILD vertex detector}
\label{fig:mechanical}
\end{figure}

Figure~\ref{fig:mechanical} shows the overall mechanical
design of the ILD vertex detector. The design is based 
on the SLD vertex detector design, but double sided ladder 
is assumed for ILD. 
Each ladder consists of 50~$\mu$m thick silicon sensors
and readout electronics chips put on both sides of ultra 
light 2-mm thick substrate made of carbon foam  or 
silicon carbide foam.
Material budget of a ladder is 0.3\%/ladder$=$0.15\%/layer.
Flexible printed circuit (FPC) cables coming from
ladders are connected to junction circuits outside the
cryostat using micro connectors. 
The junction circuit including
clock drivers, data suppression circuits, optical
fiber drivers, etc. are placed surrounding the beam pipe.
Compared with the old model used
for the simulation study in ILD Letter of Intent (LOI)~\cite{abe},
material budget in forward region is increased. 

\section{Summary}
So far, starting from 24~$\mu$m-pixel fully depleted CCD, 
small (6~mm$\times$6~mm) FPCCD prototypes have been manufactured 
four times, and finally 6~$\mu$m-pixel FPCCD has been developed.
Design of the first large prototype of FPCCD is in progress
and the prototype chips will be delivered in JFY2012.
Technology to make CCD wafers thinned down to 50~$\mu$m 
has been established. R\&D for cooling system using 2-phase
CO$_2$ is being carried out in collaboration with LC-TPC, Belle-II
vertex, and KEK cryogenic groups.
A tentative common mechanical design has been made to be 
implemented into the ILD simulation model.
 
\section*{Acknowledgments}
We would like to thank the members of CO2 collaboration
of KEK Detector Technology Project for R\&D of the 
2-phase CO$_2$ cooling system.
This work is partly supported by Grant-in-Aid for Creative 
Research No.18GS0202 and Grant-in-Aid for Specially
Promoted Research No. 23000002 by Japan Society for 
Promotion of Science (JSPS). 
This work is also partially supported by KEK 
Detector Technology Project.
 
% ****************************************************************************
% BIBLIOGRAPHY AREA
% ****************************************************************************
\begin{footnotesize}
% IF YOU DO NOT USE BIBTEX, USE THE FOLLOWING SAMPLE SCHEME FOR THE REFERENCES
% ----------------------------------------------------------------------------

% ----------------------------------------------------------------------------
\end{footnotesize}

% ****************************************************************************
% END OF BIBLIOGRAPHY AREA
% ****************************************************************************
\end{document}